\documentstyle[12pt,epsf]{article}
\newcommand {\be} {\begin{equation}}
\newcommand {\ee} {\end{equation}}
\newcommand {\beq}{\begin{eqnarray}}
\newcommand {\eeq}{\end{eqnarray}}
\newcommand {\acs}{auto correlation sum}
\newcommand {\ccs}{cross correlation sum}

\newcommand {\e}  {$\epsilon$}

\begin{document}
\begin{center}
{\Large{\bf Studying Attractor Symmetries by Means of Cross Correlation Sums}}

\bigskip

{\large {\bf Peter Schneider} and {\bf Peter Grassberger}}\\
Physics Department, University of Wuppertal, D-42097 Wuppertal, Germany

\bigskip
\bigskip

{\bf \today}
\end{center}

\bigskip

\begin{abstract}
We use the cross correlation sum introduced recently by H. Kantz to study
symmetry properties of chaotic attractors. In particular, we apply it to 
a system of six coupled nonlinear oscillators which was shown by Kroon {\it 
et al.} to have attractors with several different symmetries, and compare 
our results with those obtained by ``detectives" in the sense of 
Golubitsky {\it et al.}. 
\end{abstract}

\vspace{5.cm}

\newpage
\section{Introduction}

Spontaneous symmetry breaking is one of the most interesting problems in 
statistical physics. While in the past most work has been done on 
equilibrium systems, it has become clear only recently that very interesting 
similar phenomena occur also in non-equilibrium systems. Indeed, the variety 
of phenomena observed there is even larger than in equilibrium models, 
even if we restrict ourselves to purely deterministic (but in general 
chaotic) systems as we shall do in the present work. Spontaneous symmetry 
breaking, in particular, can be of two fundamentally different kinds. 

Assume that the equation of motion 
\be
   {d {\bf x(t)}\over dt} = {\bf F}({\bf x}(t)) 
\ee
(with ${\bf x(t)}$ in some manifold $\cal{M}$) is {\it equivariant} (or
``covariant" in the physics literature) under some symmetry group $G$ 
acting on $\cal{M}$, i.e.
\be
   g {\bf F}({\bf x}) = {\bf F}(g{\bf x})
\ee
for each element $g\in G$. We assume that ${\bf F}$ has at least one 
chaotic or quasiperiodic attractor. 

In such a case, orbits ${\bf x}(t)$ can have the following symmetry 
properties: 
\begin{itemize} 
\item (1) They are symmetric under each element $\in G$, 
i.e. $g {\bf x}(t) = {\bf x}(t)$, for each $g\in G$.
\item (2) Individual orbits are symmetric only with respect to some 
subgroup $\Sigma\subset G$, but the attractor and the invariant measure are 
still symmetric under $G$. Notice that $\Sigma$ 
can be trivial (having just the identity as unique element), in which case 
the {\it instantaneous} symmetry is maximally broken, while the symmetry 
of the attractor is still unbroken. Such a case occurs most naturally when 
the system is composed of two identical subunits and $G$ is just the 
permutation group of two elements interchanging them. The transition from 
item 2 to item 1 manifests itself then in synchronization of the two subunits 
\cite{fuji,pikov,pikov-grass}. Thus (de-)synchronization is just the simplest 
version of a sort of spontaneous symmetry breaking, and it was stressed in
\cite{grass-pikov} that the breaking of other symmetries should follow 
essentially the same lines.   
\item (3) Also the attractor is no longer symmetric under the full group $G$, 
but only under some subgroup $\Sigma\subset G$ (the ``symmetry group" 
of the attractor). In this case one has of 
course coexistence of several attractors related to each other by symmetry 
transformations $g\in G/\Sigma$. Transitions between item 1 and item 3
are possible \cite{ashwin} but less common than transitions either 
between 1 and 2 or between 2 and 3. 
The latter have been studied intensively in \cite{chossat,field,king-stew}.
In the most frequent scenario, the different attractors in the symmetry 
broken phase are disjoint, and symmetry is restored when these attractors 
touch in one or several points.
\end{itemize}

In the present paper we shall only deal with item 3, and with transitions 
between 2 and 3. We shall not deal with item 1 or with transitions 
between 1 and 2. More precisely, we 
shall study the following problem \cite{barany,dellnitz,kroon}: we are 
given a time series generated by ${\bf F}$, and we want to estimate the 
maximal subgroup $\Sigma\in G$ under which the natural invariant measure 
({\it BRS-measure} \cite{eck-ruelle}) is invariant, 
\be 
   \mu(B) = \mu(h^{-1}B)\;,\qquad h\in \Sigma \;.
\ee  
Here, $B$ is any measurable set intersecting the attractor, and we 
assume that the initial conditions for the time series are $\mu$-random. 
Since this is 
a statistical decision problem, we cannot expect an unambiguous answer. 
In a proper treatment we should work out the detailed statistics of 
any used observable, and obtain the corresponding significance criteria. 
Following \cite{barany,dellnitz,kroon} and a venerable tradition 
established in physics, we shall not do so here but hope that the results 
will be so clear cut that such formal significance criteria will not be 
necessary.

In order to find the symmetry of a measure $\mu$ from a set of $N$ points 
${\bf x}_k$, $k=1\ldots N$ drawn randomly from this measure, there are 
several possible strategies. 

(a) Visual inspection. For low dimensions, this is of course the method of 
choice. At higher dimensions, we can still visualize by making projections 
or cross sections. But usually, these will be less informative and more 
cumbersome.

(b) We can partition phase space in some way such that each $g\in G$ acts 
as a permutation $P_g$ on the elements ${\cal U}_i$ of the partition, 
\be
   g \;{\cal U}_i = {\cal U}_j\;, \qquad j = P_g i \;.
\ee 
If the measure is symmetric, then we have
\be 
   \mu({\cal U}_i) = \mu({\cal U}_{P_g i}) .
\ee
Therefore, if the time series points are iid (identically and 
independently distributed), then
\be 
   n_i \approx n_{P_g i} \pm O(\sqrt{n_i}) , 
\ee
where $n_i$ is the number of points in ${\cal U}_i$. In simple 
cases this can be very efficient, but it becomes very unwieldy in 
high dimensions and if $G$-invariant partitions are tricky to construct.

(c) We can consider averages over functions which themselves are not 
invariant under $G$. Assume that $\phi({\bf x})$ is such a function. Then 
\be
   \langle \phi \rangle  \equiv N^{-1}\sum_{k=1}^N \phi({\bf x}_k)
\ee
will in general be different from
\be
   \langle g\phi \rangle  =  N^{-1}\sum_{k=1}^N \phi(g^{-1}{\bf x}_k)
\ee
unless $\mu$ is invariant under $g$. Functions $\phi$ for which 
$\langle \phi \rangle = \langle g\phi \rangle$ {\it only} if $\mu$ is 
invariant are called {\it detectives} in \cite{barany,dellnitz,kroon} 
(see these references for a more precise definition), and the numerical 
study of symmetry properties in these papers is based mainly on this 
method and on visual inspection of projections.

(d) In the present paper, we shall use yet another method based on an 
idea by H. Kantz \cite{kantz}. In that work, the author was concerned with 
the more general problem of verifying numerically whether two measures 
are identical, making use of two random samples of points drawn from these 
measures.  It uses the {\it cross correlation sum}, a generalization of the 
correlation sum used e.g. in estimates of the correlation dimension.
 
Assume that the two sets of points are  $X = \{{\bf x}_k, \; k=1\ldots N\}$ 
and $Y = \{{\bf y}_k, \; k=1\ldots N\}$. They are randomly drawn from the 
same space, but according to two different measures $\mu$ and $\nu$. The 
correlation sum for the set $X$ is defined as 
\be 
   C_{XX}(\epsilon)=\frac{1}{N(N-1)} \sum_{{\bf x,y}\in X;\;{\bf x}\neq{\bf y}} 
      \Theta(\epsilon-|{\bf x-y}|)   \label{corr}
\ee
where $\Theta$ is the Heavyside step function. Thus $C_{XX}$ measures the 
fraction of pairs having a distance $<\epsilon$.
The cross correlation sum is defined similarly by 
\be
   C_{XY}(\epsilon)=\frac{1}{N^2} \sum_{{\bf x}\in X} \sum_{{\bf y}\in Y} 
      \Theta(\epsilon-|{\bf x}-{\bf y}|) \;.   \label{c_corr}
\ee
Obviously, $C_{XY}\approx C_{XX}$ up to statistical fluctuations if $X$ and 
$Y$ are distributed according to the same measure. Otherwise, we expect an 
inequality. More precisely, let us define the cross correlation {\it 
integral} by
\be
    C_{\mu\nu}(\epsilon) = \int d\mu({\bf x})\;\int d\nu({\bf y})\; 
              \Theta(\epsilon-|{\bf x-y}|)\;.
\ee
Almost surely, $C_{XY}(\epsilon) \to C_{\mu\nu}(\epsilon)$ for 
$N\to\infty$, and $C_{XY}(\epsilon)$ is an unbiased estimator for 
$C_{\mu\nu}(\epsilon)$ in the sense that $C_{\mu\nu}(\epsilon) = 
\langle C_{XY}(\epsilon) \rangle$ where the brackets denote the average 
over different random samples $X$ and $Y$. For sufficiently small
\e\ and for absolutely continuous measures $\mu$ and $\nu$ one can then
show \cite{kantz}
\be
   C_{\mu\nu}^2(\epsilon) \; \leq\; C_{\mu\mu}(\epsilon) \;
                                    C_{\nu\nu}(\epsilon) \;. \label{ineq}
\ee
Unfortunately, the same is not necessarily true for $C_{XY}(\epsilon)$
with finite $N$ \cite{kantz}, due to the omission of ``diagonal" terms in
the double sum in eq.(\ref{corr}). This omission is necessary since
otherwise $C_{XX}(\epsilon)$ would be a biased estimator for
$C_{\mu\mu}(\epsilon)$. Nevertheless, the analogon of eq.(\ref{ineq}) for
finite $N$ will be violated only in exceptional cases due to
statistical fluctuations, and only if $\mu\approx \nu$.

Also, eq.(\ref{ineq}) cannot be shown for all \e\ and for arbitrary
measures. This follows from Bochner's theorem and the fact that the
Fourier transform of $\Theta(\epsilon-|{\bf x}|)$ is not positive definite.
To avoid this problem, we could e.g. replace $\Theta$ by a Gaussian,
so that $
    C_{\mu\nu}(\epsilon) = \int d\mu({\bf x})\;\int d\nu({\bf y})\;
              e^{-({\bf x-y})^2/\epsilon}\;$
and similarly for the discrete cross correlation sums.
We did not do this since this would make the numerics less efficient,
and since we did not encounter any practical problems when using
eq.(\ref{c_corr}) in the applications described in sec.3.
                                                                      
Thus we use as a similarity measure the cross correlation ratio
\be
   r(\epsilon) = \frac {C_{XY}(\epsilon)}{\sqrt{C_{XX}(\epsilon)
                  C_{YY}(\epsilon)}}\;.
\ee
For very large $\epsilon$ it is equal to 1 since $C_{XY}(\epsilon) = 
C_{XX}(\epsilon)=C_{YY}(\epsilon)=1$ for sufficiently large $\epsilon$ 
(we assume that the measures have compact support). On the other extreme, 
$r(\epsilon)\to 0$ for $\epsilon \to 0$ if the measures $\mu$ and $\nu$ 
have disjoint supports. The latter is typically \cite{chossat} but not 
always \cite{ashwin} true if $\mu$ is a (time) invariant measure which 
is not invariant under some $g\in G$, and $\nu$ is the $g$-image of $\mu$. 
For intermediate values of $\epsilon$ we expect a monotonic increase, 
though we cannot exclude a more complex behavior. We can define a 
characteristic distance $\epsilon_c$ between the sets $X$ and $Y$ by 
demanding for instance $r(\epsilon_c)=1/2$.

In the present application, $X$ is a time series and $Y$ is just the image 
of $X$ under a group action $g$, 
\be
   Y = gX \equiv \{g{\bf x} |\; {\bf x}\in X\} \;.
\ee
Furthermore, we assume that the norm $|.|$ is invariant under $g$ (in 
the following, we use Euclidean norm). Thus $C_{YY}=C_{XX}$, and the 
cross correlation ratio for element $g$ is equal to 
\be
   r_g(\epsilon) = \frac {C_{X,gX}(\epsilon)}{C_{XX}(\epsilon)} \;.
\ee

A problem arises from the fact that points in a time series are correlated 
and do thus not form a sample of iid points. This has mainly two 
more or less disjoint effects. 
The first is that it might take rather long until an orbit visits all 
parts of the attractor. This is particularly important in the vicinity 
of a bifurcation, and when there is intermittency or transients with 
very long characteristic times $\tau$. In such cases the only remedy is 
to take time sequences which span a total time $T>>\tau$. To avoid huge 
data sets, one should then decimate the data. We thus take 
\be 
   {\bf x}_n = {\bf x}(nm\;dt) \;,\quad n=1,\ldots N      \label{decimate}
\ee
where $dt$ is the integration step and $m$ is a sufficiently large integer. 
Independent from this, data on short scales might be correlated on time 
scales $t<t_{corr}$ with $t_{corr} >m\;dt$. In this case, even the 
decimated data are correlated. As pointed out by Theiler \cite{theiler}, 
it is then necessary to omit pairs in eqs.(\ref{corr},\ref{c_corr}) whose 
indices differ by less then $n_0 = t_{corr} /m\;dt$. Thus our final 
formula for $r_g(\epsilon)$ is 
\be
   r_g(\epsilon) = \frac {\sum\sum_{i,k=1}^N 
        \Theta(\epsilon-|{\bf x}_i-g{\bf x}_k|)\,\Theta_{|i-k|\geq n_0}}
                        {{\sum\sum_{i,k=1}^N
        \Theta(\epsilon-|{\bf x}_i-{\bf x}_k|)}\,\Theta_{|i-k|\geq n_0}}\;.  
\ee   
Finally let us discuss some immediate consequences following from the facts 
that $C_{XY}$ is symmetric under the exchange $X\leftrightarrow Y$, that 
$G$ is a group, and that the norm is invariant under $G$. The first is 
that $C_{X,gX}=C_{gX,X}=C_{X,g^{-1}X}$ for all $g\in G$, and hence 
\be
   r_g(\epsilon)=r_{g^{-1}}(\epsilon)\;,\quad g\in G\,.    \label{r-ident}
\ee
Next assume that $h$ is in the symmetry group $\Sigma$ of the attractor. 
Then $hX\approx X$ and thus $C_{X,gX} \approx C_{X,ghX}\approx C_{X,hgX}$ 
with errors vanishing for $N\to\infty$. This implies that $r_g$ takes the 
same value in the entire left and right cosets of $\Sigma$,
\be
   r_g(\epsilon)=r_{gh}(\epsilon)=r_{hg}(\epsilon)\;,\quad h\in \Sigma\;
                         ,\;\;  g\in G\,,            \label{r-coset}
\ee
up to statistical fluctuations. 

\section{Numerical Example}

Since we want to test our method against the method of detectives 
\cite{barany,dellnitz,kroon}, we use exactly the same model as in 
\cite{kroon}. It models the gait of a hexapodal animal, and is described 
in detail in \cite{collins}. It involves six identical nonlinear 
oscillators, coupled in a ring-like fashion. More precisely, the equations 
of motion are (note the misprint in the second line in \cite{kroon})
\begin{eqnarray}
   \frac{dx_p}{dt}&=&-4x_p+y_p+(x_p^2+y_p^2)(Px_p-Qy_p)- \cr
      & -&\lambda(4(x_{p-1}+x_{p+1})-2(y_{p-1}+y_{p+1}))\\  \nonumber
   \frac{dy_p}{dt}&=& -x_p-4y_p+(x_p^2+y_p^2)(Py_p+Qx_p)- \cr
      & -&\lambda(2(x_{p-1}+x_{p+1})+4(y_{p-1}+y_{p+1}))\;,   \label{dg2}
       \;\quad p=0,1,\ldots 5\; ({\rm mod}\; 6)\;.
\end{eqnarray}
The system is equivariant under cyclic or anticyclic permutations
of the indices of the oscillators. This would give the dihedral group 
${\bf D}_6$\ as equivariance group. It consists of 12 
permutations and is a subgroup of ${\bf S}_{12}$. Its generators are
\be
  \kappa=\left( \begin{array}{cccccc} 0 & 1 & 2 & 3 & 4 & 5\\
    0 & 5 & 4 & 3 & 2 & 1 \end{array} \right)
\ee
\be
  \zeta =\left( \begin{array}{cccccc} 0 & 1 & 2 & 3 & 4 & 5\\
    1 & 2 & 3 & 4 & 5 & 0 \end{array} \right) \;.
\ee

But in addition, the above system has a ${\bf Z}_2$\ symmetry 
$(x_p,y_p)\to (-x_p,-y_p)$ which was missed in \cite{kroon}. 
Thus the full equivariance group is 
\be 
    G = {\bf D}_6 \times {\bf Z}_2
\ee

Attractors can have as symmetry groups any subgroup of $G$. 
Disregarding the factor ${\bf Z}_2$ for the moment, the subgroups 
of ${\bf D}_6$ are (up to conjugacies) {\bf 1}, ${\bf Z}_2 =\{1,
\zeta^3\}$, ${\bf Z}_3$, ${\bf Z}_6$, ${\bf D}_1=\{1,\kappa\}$, 
$\tilde{\bf D}_1=\{1,\kappa\zeta\}$, ${\bf D}_2$, ${\bf D}_3$, 
$\tilde{\bf D}_3$ and ${\bf D}_6$.

In \cite{kroon}, the attractor was projected for visual inspection 
onto a two dimensional subspace by means of the variables 
\be
   u = x_1-x_2+x_4-x_5       \label{u}
\ee
and 
\be
   v = {1\over \sqrt{3}} (x_1+x_2-2x_3+x_4+x_5-2x_0) \;.
                             \label{v}
\ee
In the $(u,v)$ plane, the action of $\zeta$ is a clockwise 
(negative) rotation by an angle $2\pi/3$, while $\kappa$ 
corresponds to a reflection on the axis $u=0$. Notice that this 
is not a faithful representation of ${\bf D}_6$ since $\zeta^3$ is 
represented by the unity $(u,v)\to(u,v)$. The figures shown in 
\cite{kroon} suggest that all attractors are symmetric under the 
additional ${\bf Z}_2$ symmetry. This was also verified by ourselves 
in all cases, whence we shall disregard this ${\bf Z}_2$ in the 
following and consider only symmetry under the elements of 
${\bf D}_6$.

As distance measure in the correlation sums we used 
$r = [\sum_p(x_p-x'_p)^2 + \sum_p(y_p-y'_p)^2]^{1/2}$.

Since our goal was to compare the correlation sum method to the 
method of detectives, we followed \cite{kroon} in keeping the
parameters $P$ and $\lambda$ fixed at $P=-5$ and $\lambda=1.2$. 
The parameter $Q$ was used as control parameter and was varied 
in the range $25.0 \leq Q \leq 27.2$ studied also in \cite{kroon}.

We integrated eq.(\ref{dg2}) with a fourth order Runge-Kutta 
algorithm with time step $dt=0.01$. Again this was done in 
order to compare with \cite{kroon}. This 
time step might be too large for some of the $Q$-values 
considered, whence some of the observed structures might be 
artefacts of the time discretization. But this is not much of a 
concern: the time discretized system is as good a nonlinear 
dynamical system as the ODE, and studying its symmetries should 
be neither more easy nor more difficult. 

An important point where we departed from \cite{kroon} was the 
global time scales. We found that several of the attractors described 
in \cite{kroon} represented actually transients (e.g. for $Q=26.9$, 
the final attractor is as shown in fig.3 below, and not as 
described in \cite{kroon}). Thus we discarded in all cases a 
transient of length $t\geq 10000$. In addition, we have the problem 
of long residence times on parts of the entire attractor discussed 
already in \cite{kroon}. This gives rise to fake attractor 
asymmetries if the length of the time series is less than these 
residence times. To avoid this, we took only every 550th 
point\footnote{For 
quasiperiodic attractors we found that large spurious correlations 
with very slow temporal decay could arise from near commensurability 
between the sampling frequency and some internal frequency of the 
attractor. To avoid this, we actually used an irregular sampling 
where we selected $m$ randomly between 100 and 1000,
such that $\langle m \rangle = 550$. Without this trick,
we would have obtained $r_g(\epsilon)>>1$ in several cases.}, 
i.e. $m\;dt \approx 5$ (see eq.(\ref{decimate})).
Finally, we used a Theiler cutoff $n_0=100$. 

Another point where we departed from \cite{kroon} was the 
number of points used. While the analysis of \cite{kroon} was 
based on 10,000 
- 50,000 points for each value of $Q$, we used only 5,000 points. 
This was done not only for minimizing the computational effort
(which is higher for our method), but also to demonstrate that 
our method is very efficient already for small data samples.

A last remark: as one should expect in general, we found in some 
cases that the system possessed several attractors which are not 
related by symmetry. For instance, depending on the initial 
conditions we found for $Q=25.7$ either the attractor shown in 
\cite{kroon} or another one which closely resembles the one found 
in \cite{kroon} at $Q=25.6$.

\section{Results}

In figures 1 to 4 we show four typical attractors together with 
the corresponding (cross-) correlation sums. For visualizing the 
attractors, we use the projection given by eqs.(\ref{u}),(\ref{v}). 
In each correlation sum plot, the number of pairs closer than \e\ 
are plotted against \e\, for each of the ten group elements $g = 
1,\,\zeta,\, \zeta^2,\, \zeta^3,\, \kappa,\, \kappa\zeta,\, 
\kappa\zeta^2,\, \kappa\zeta^3$, $\kappa\zeta^4$ and 
$\kappa\zeta^5$. Results for the remaining elements 
\begin{figure}[htb]
\unitlength1cm
\hfill \\
\begin{minipage}[t]{5.0cm}
\mbox{  \\         }
\begin{picture}(5.0,5.0)
\centerline{\epsfbox{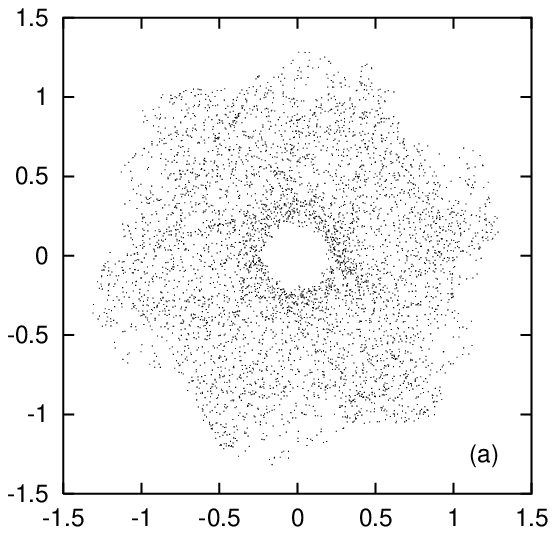}}
\end{picture}\par
\end{minipage}
\hfill
\begin{minipage}[t]{7.0cm}
\mbox{  \\         }
\begin{picture}(7.0,5.0)
\centerline{\epsfbox{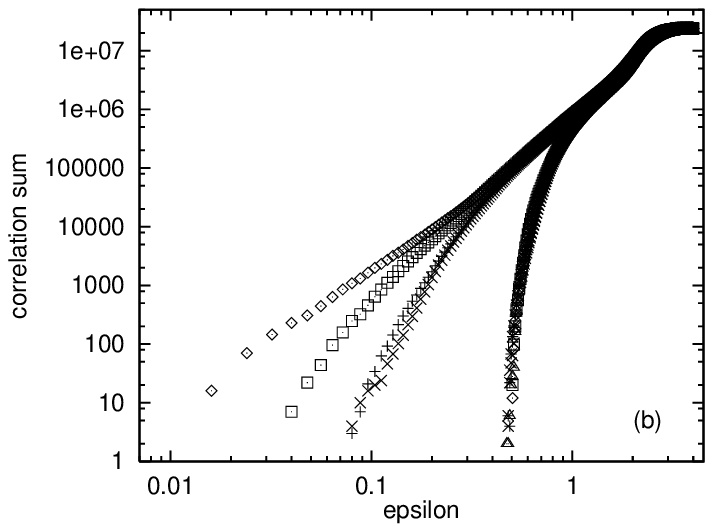}}
\end{picture}\par
\end{minipage}
\caption{\small Panel (a): attractor for $Q=25.04$ projected onto the
$(u,v)$-plane by means of eqs.(\ref{u},\ref{v}). Panel (b):
log-log plot of the cross and auto correlation
sums (without normalization factors $N^{-2}$) for $Q=25.04$.
The different symbols have the following meaning: $\Diamond$: $g=1$
or $\kappa\zeta^2$; +: $g=\zeta$ or $\kappa\zeta^3$; $\Box$: $g=\zeta^2$ 
or $\kappa\zeta^4$; $\times$: $g=\zeta^3$ or $\kappa\zeta^5$; 
$\triangle$: $g=\kappa$; $\ast$: $g=\kappa\zeta$.}
\end{figure}
\begin{figure}[htb]
\unitlength1cm
\begin{minipage}[t]{5.0cm}
\mbox{  \\         }
\begin{picture}(5.0,5.0)
\centerline{\epsfbox{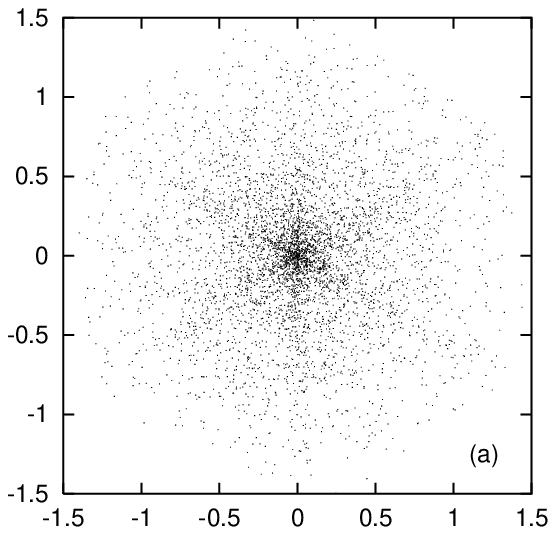}}
\end{picture}\par
\end{minipage}
\hfill
\begin{minipage}[t]{7.0cm}
\mbox{  \\         }
\begin{picture}(7.0,5.0)
\centerline{\epsfbox{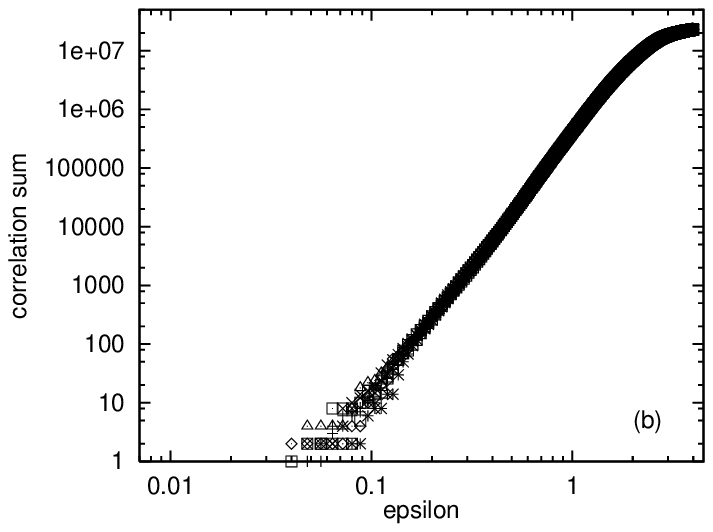}}
\end{picture}\par
\end{minipage}
\caption{\small Same as fig.1, but for $Q=27.2$.}
\end{figure}
$\zeta^4$ and $\zeta^5$ are not shown since they coincide due 
to eq.(\ref{r-ident}) identically with those for $\zeta^2$ and 
$\zeta$, respectively. Also, as we said already in the last 
section, we do not show the cross correlations for the inversion 
$({\bf x},{\bf y}) \to (-{\bf x},-{\bf y})$ since all attractors 
were found to be symmetric under it. We show the 
numbers of pairs instead of the correlation sums with 
conventional normalization as in eq.(\ref{corr}), since the 
square roots of these numbers give rough estimates of statistical 
errors for small values of \e\ where these errors are most 
important (for rigorous error estimates of autocorrelation 
sums, see \cite{bds,wu}).

If an attractor is symmetric with respect to a group element $g$, 
we expect the corresponding $C_{X,gX}$ to coincide with $C_{X,X}$. 
Within the expected statistical errors, the symmetries found in this 
way were also verified visually in the projections, except for 
$g=\zeta^3$ which is mapped by the projection (\ref{u}),(\ref{v}) 
onto the unit element. Thus invariance under $\zeta^3$ cannot be read 
off from figures 1a to 4a, and can be seen only from the corresponding 
correlation sum plots.

\begin{figure}[htb]
\unitlength1cm
\hfill \\
\begin{minipage}[t]{5.0cm}
\mbox{  \\         }
\begin{picture}(5.0,5.0)
\centerline{\epsfbox{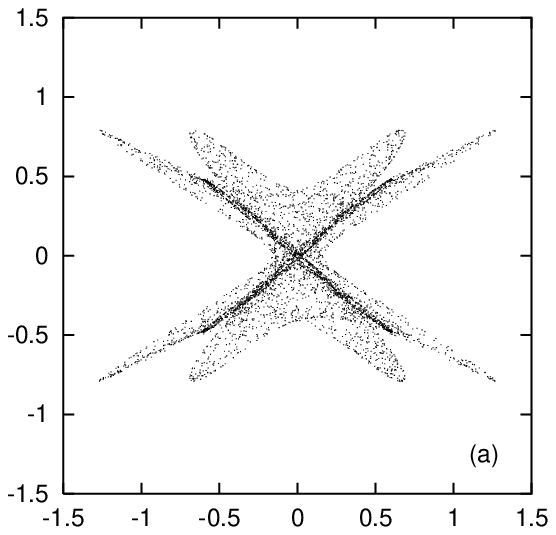}}
\end{picture}\par
\end{minipage}
\hfill 
\begin{minipage}[t]{7.0cm}
\mbox{  \\         }
\begin{picture}(7.0,5.0)
\centerline{\epsfbox{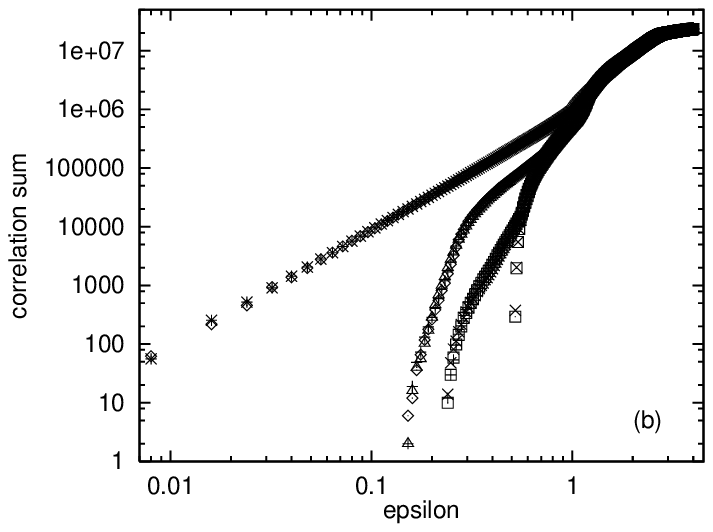}}
\end{picture}\par
\end{minipage}
\caption{\small Same as fig.1, but for $Q=26.9$.}
\end{figure}
\begin{figure}[htb]
\unitlength1cm
\begin{minipage}[t]{5.0cm}
\begin{picture}(5.0,5.0)
\centerline{\epsfbox{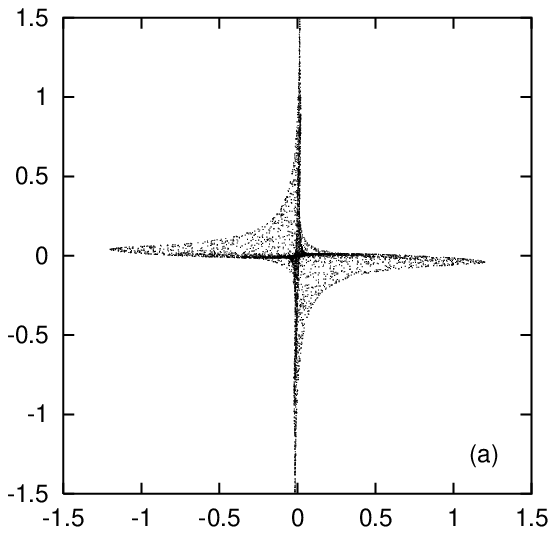}}
\end{picture}\par
\end{minipage}
\hfill
\begin{minipage}[t]{7.0cm}
\begin{picture}(7.0,5.0)
\centerline{\epsfbox{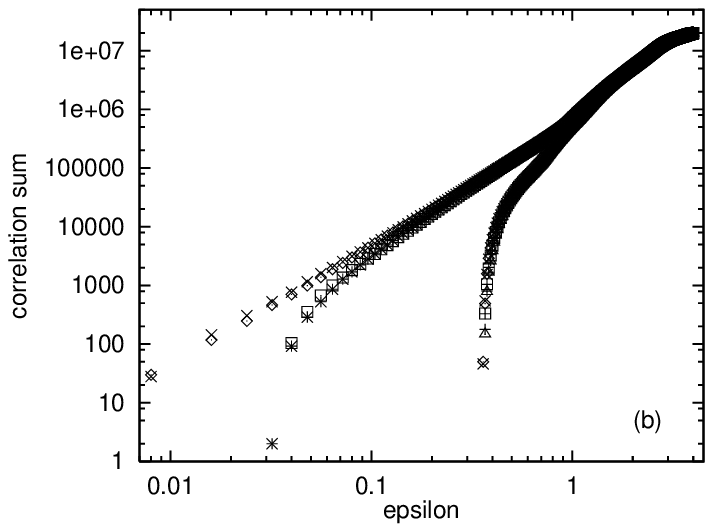}}
\end{picture}\par
\end{minipage}
\caption{\small Same as fig.1, but for $Q=26.1$.}
\end{figure}

In fig.1, the projection suggests a symmetry under the inversion
$(u,v)\to (-u,-v)$, and maybe even under a rotation by $60^0$.
We might be tempted to interpret this as invariance of the
original attractor under ${\bf Z}_2$ or ${\bf Z}_6$. From the
correlation sums we see that this is not correct. Apart from the
inversion $({\bf x},{\bf y}) \to (-{\bf x},-{\bf y})$,
the attractor has no symmetry at all, since none of the
cross correlations agrees with the \acs\ (we notice that the
invariance under $(u,v)\to (-u,-v)$ seen in all figures in
\cite{kroon} is due to this inversion, and is not due to the
particular projection used).

The other extreme of a completely symmetric attractor is shown
in fig.2. We see that all \ccs s coincide.

An example of $\tilde{\bf D}_1$ invariance is shown in fig.3. In
the correlation sum plot we see four different curves. The one on
top contains $C_{X,X}$ and $C_{X,\kappa\zeta X}$. The other
curves contain $C_{X,\zeta X},\, C_{X,\kappa X},\, C_{X,\kappa
\zeta^2 X}$ (second from top), $C_{X,\zeta^2 X}$, $C_{X,\kappa
\zeta^3 X},\, C_{X,\kappa \zeta^5 X}$ (third from top), and
$C_{X,\zeta^3 X},\, C_{X,\kappa \zeta^4 X}$ (bottom/rightmost
curve). This degeneracy of curves follows immediately from
eq.(\ref{r-coset}) and the detailed group structure of ${\bf D}_6$.

Finally, we show in fig.4 a case with ${\bf Z}_2$ symmetry. We
seem to see there three curves, but the rightmost one indeed
consists of a triplet of lines which can only be resolved at higher
resolution. Thus we have five curves altogether, each corresponding
to two \ccs s. This is exactly as expected from eq.(\ref{r-coset}).
The near degeneracy of the rightmost three lines is explained
by the very weak breaking of ${\bf D}_2$ symmetry which is also
seen in the fact that the two top curves are not very different
except for very small \e .

\begin{figure}[p]
\centerline{\epsfbox{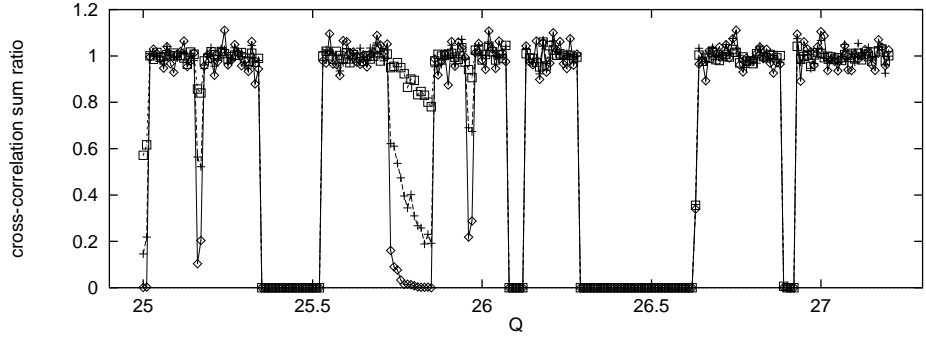}}
\caption{\small Cross correlation coefficients $r_g(\epsilon)$ for
$g=\zeta$, plotted against $Q$. Three data sets are superimposed,
for distances \e\ at which 1000, 3000 and 10,000 $\epsilon$-close
pairs were found. The corresponding symbols are $\Diamond$ (1000),
+ (3000), and $\Box$ (10,000).}
\end{figure}
\begin{figure}[p]
\centerline{\epsfbox{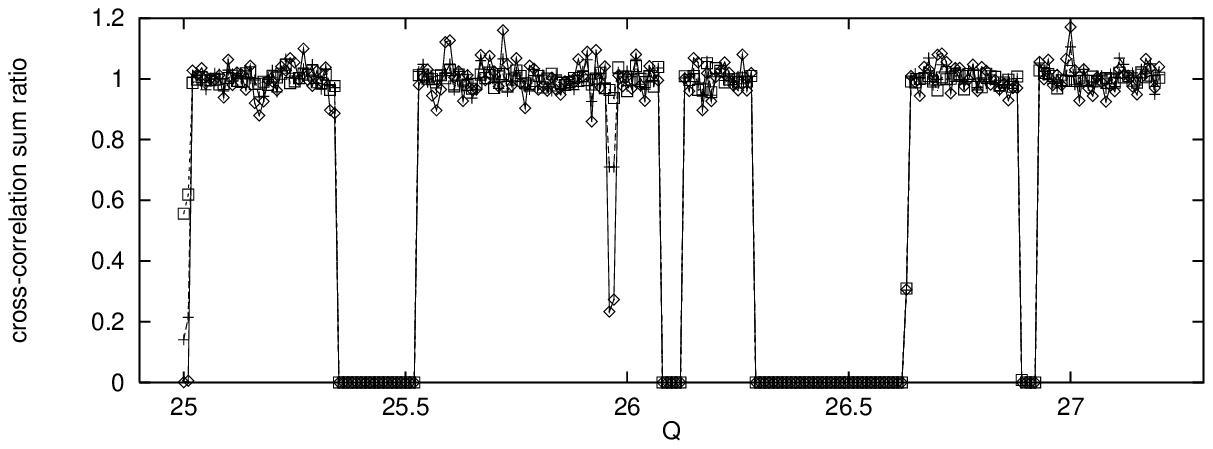}}
\caption{\small Same as fig.5, but for $g=\zeta^2$.}
\end{figure}
\begin{figure}[p]
\centerline{\epsfbox{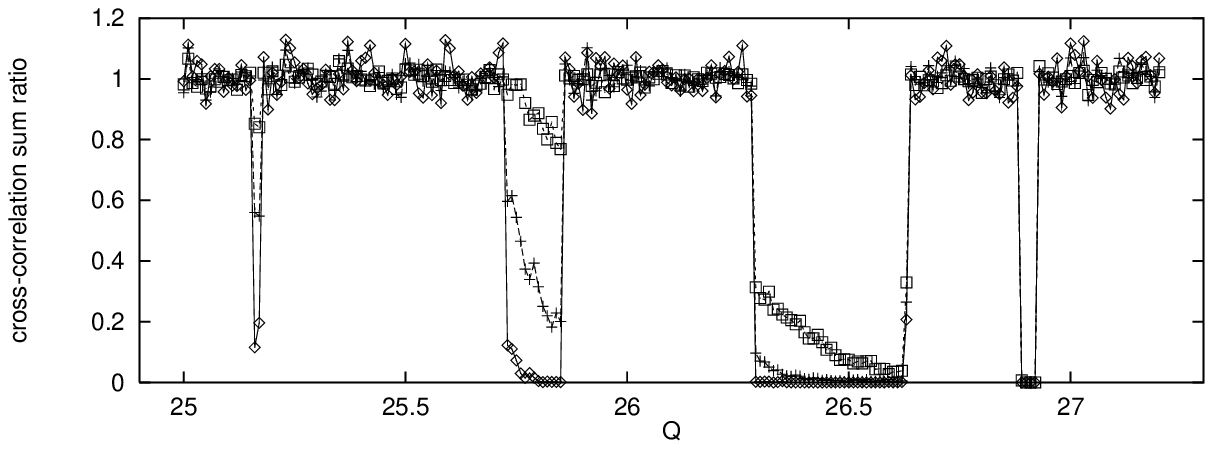}}
\caption{\small Same as fig.5, but for $g=\zeta^3$.}
\end{figure}
\begin{figure}[p]
\centerline{\epsfbox{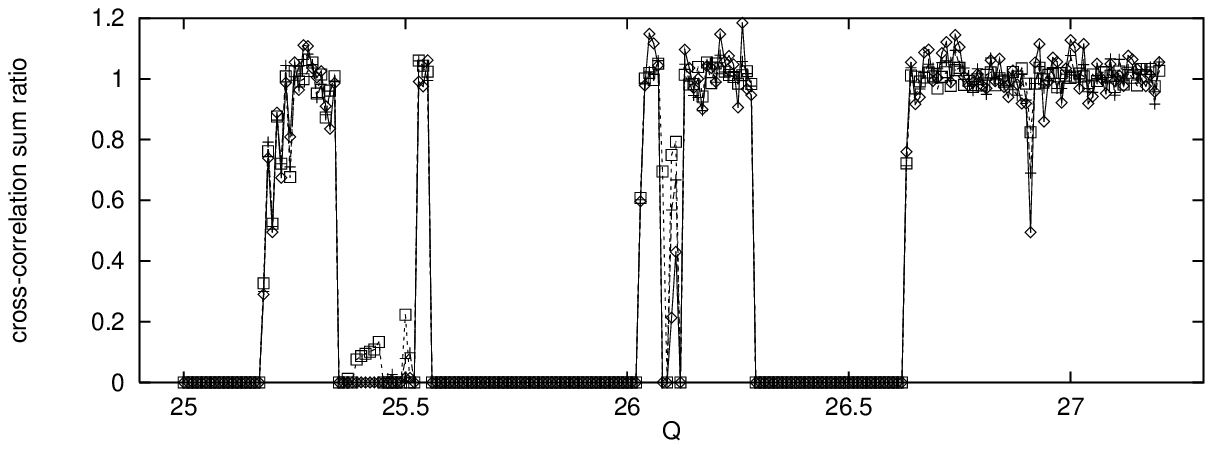}}
\caption{\small Same as fig.5, but for $g=\kappa$.}
\end{figure}
\begin{figure}[p]
\centerline{\epsfbox{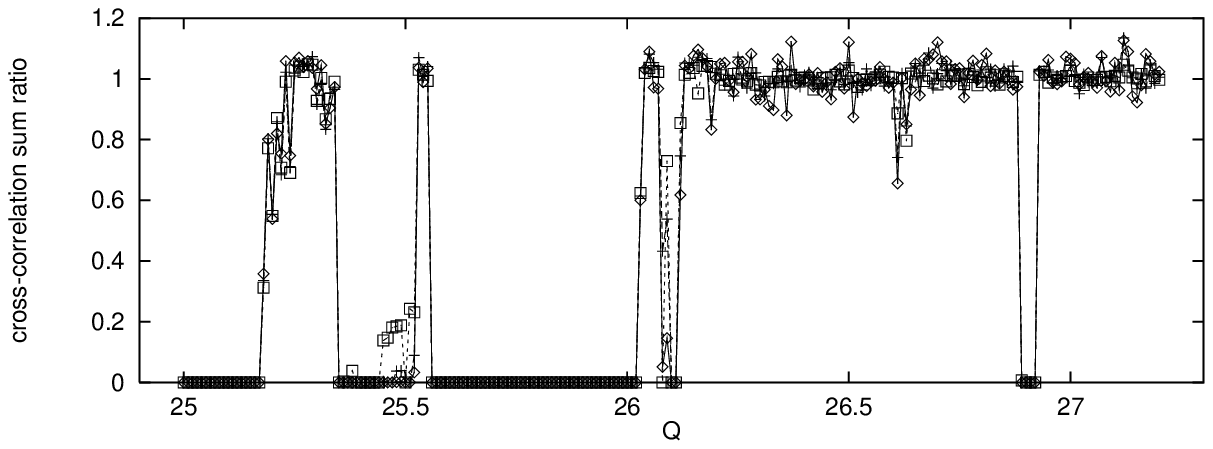}}
\caption{ Same as fig.5, but for $g=\kappa\zeta$.}
\end{figure}
\begin{figure}[p]
\centerline{\epsfbox{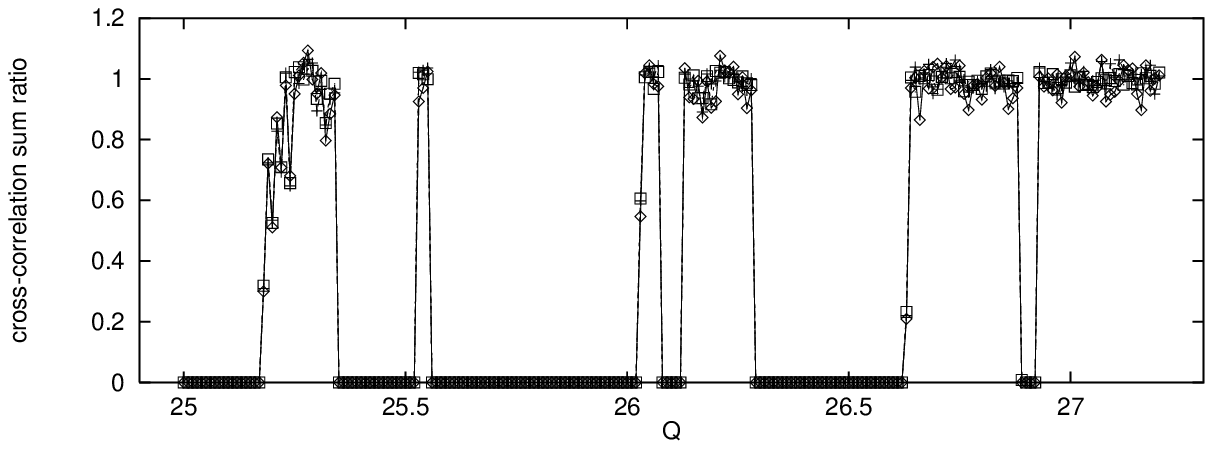}}
\caption{\small Same as fig.5, but for $g=\kappa\zeta^2$.}
\end{figure}

In all examples (including many more which are not shown here), 
the symmetry suggested by the correlation sums is compatible 
with that seen by looking at the $(u,v)$-projection of the 
attractor. 

For a systematic investigation we have repeated this analysis
for all values of $Q\in [25.0,\,27.2]$ in steps of $\Delta Q =
0.01$ (221 values alltogether). We repeated this three times
with different types of initial conditions, in order to test
for multiple attractors. In the first set of runs, we used the
same initial values for all $Q$. In the second, different
random initial conditions were used for each $Q$. Finally, in
the third set of runs we used the last configuration from the
simulation with $Q$ as initial condition for simulating at
$Q+\Delta Q$. In some cases (such as $Q=2.57$, e.g.) this
gave clear indications of multiple attractors which are not
simply rotated images of each other.
\begin{figure}[h]
\centerline{\epsfbox{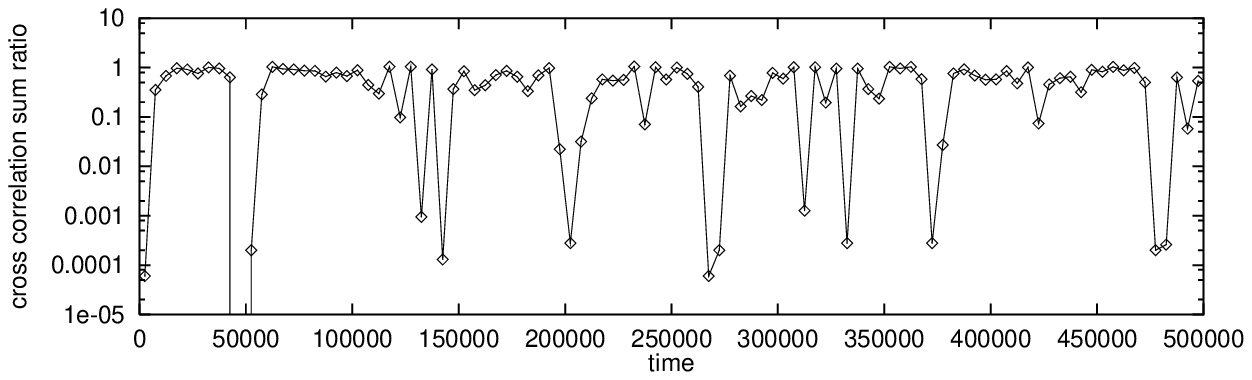}}
\caption{\small Cross correlation coefficient for $g=\kappa$,
$Q=25.22$ and $\epsilon=0.36$ plotted against time. For this,
a very long time series was cut into pieces  each containing
5000 points, and correlation coefficients were computed for
each piece independently. On the horizontal axis is plotted
physical time. In contrast to the previous plots, the average
sampling frequency is $\approx 1$.}
\end{figure}
\begin{figure}[h]
\centerline{\epsfbox{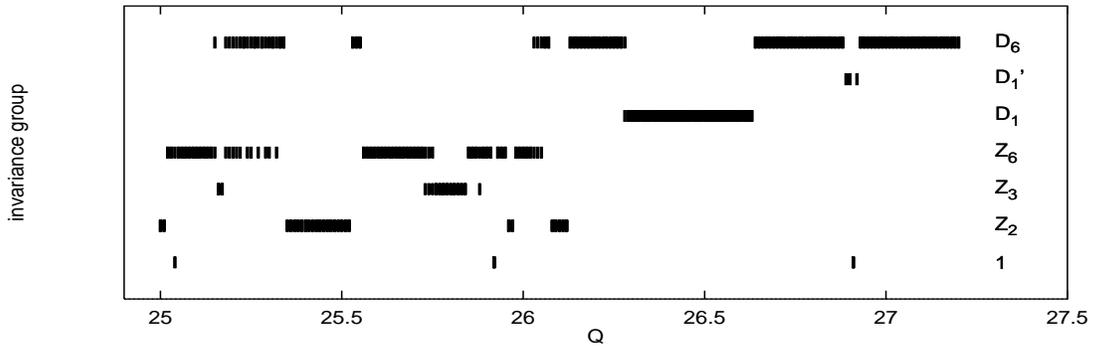}}
\caption{\small Final symmetry attributions versus $Q$. Notice that
symmetries ${\bf D}_2, {\bf D}_3$ and $\tilde{\bf D}_3$ never appear. }
\end{figure}

Plots of cross correlation sum ratios against $Q$ are shown in 
figures 5-10. These plots show results from the third type of 
initial conditions\footnote{For simplifying the figures, each 
attractor was first brought by a conjugacy into a standard 
orientation such that the maximal $r_g$ for reflextions $g=\kappa
\zeta^n\,,\;n=0\ldots 5,$ was obtained at either $g=\kappa$ or 
$g=\kappa\zeta$.}, but the plots obtained with the other two 
types of inital conditions were very similar. Each figure
corresponds to one group element. In each figure, three 
curves are shown. They represent $r_g(\epsilon)$ at those 
three values of \e\ where the unnormalized autocorrelation sum 
was equal to 1000, 3000 and 10,000, respectively. In this 
way we guaranteed that the statistical errors were roughly 
independent of $Q$, although the size of the attractor depends 
on it considerably.

We do not show plots for $g=\kappa\zeta^3, \kappa\zeta^4$ and 
$\kappa\zeta^5$ since we found that $r_g=1$ for these elements if and 
only if $r_{\kappa\zeta^2}=1$ (within statistical errors, of course). 
Thus for all values of $Q$ the symmetry group $\Sigma_Q$ either 
contains all the elements $\kappa\zeta^2, \kappa\zeta^3, \kappa
\zeta^4$ and $\kappa\zeta^5$, or none of them. This implies that none 
of the attractors had ${\bf D}_2,\, {\bf D}_3$, or ${\bf \tilde{D}_3}$ 
as its symmetry group, since all these groups (including all their 
conjugates) contain some elements of the set $\{\kappa\zeta^2, 
\kappa\zeta^3, \kappa\zeta^4, \kappa\zeta^5\}$ but not all of them.

For most values of $Q$ this gave unique symmetry attributions, 
except for very few points mostly near $Q\approx 25.2$. In this 
range, the orbit switches occasionally between different 
geometric objects (which are of course not attractors in the 
strict sense), with vastly different time scales. This had 
already been noticed in \cite{kroon}. As an example we show in 
fig.11 the correlation sum ratio for $g=\kappa$, $\epsilon
=0.36$ and $Q=25.22$ as a function of time. For this figure, a 
very long trajectory ($t=5\times 10^5$ and $5\times 10^5$ points 
as well) was partitioned into 100 pieces, and 
correlation sums were computed for each piece seperately. We see 
how $r_{\kappa}$ switches beween 0 and 1, with a typical 
time scale of $10^4$.

If this time scale is comparable to the time span covered by the 
set $X$, the latter looks as if obtained from 
a superposition of two or more measures. The same is 
obtained if part of the time series comes from a very long 
lived repellor. It is not easy to distinguish between these two 
possibilities using correlation sums only. But the presence of 
either of them is easily recognized in figs.5-10, since 
$r_g(\epsilon)$ is roughly independent of \e\ except for very 
small and very large values of the latter. 

Our final results are shown in fig.12 where we have indicated the 
invariance group for every $Q$. Multiple entries result from 
coexisting attractors and/or from the above ambiguity.

When comparing these results with those of \cite{kroon}, we observe
several differences: \\
(a) We see much more structure simply because we use a finer 
sampling in $Q$ ($\Delta Q = 0.01$ instead of 0.05 resp. 0.1). For 
that reason some of the very narrow windows as those at $Q=25.92$ 
and at $Q=25.96-25.97$ with $\Sigma = {\bf 1}$ and ${\bf Z}_2$ were 
simply missed in \cite{kroon}. \\
(b) The window at $Q=26.89-26.92$ with symmetries ${\bf 1}$ and 
${\bf \tilde{D}}_1$ was presumably missed in \cite{kroon} since it 
becomes visible only after a very long transient has died out.
A long transient (and not intermittency as suggested in \cite{kroon}) 
is also the reason for the transition from ${\bf D}_2$ to ${\bf D}_6$ 
at $Q=26.15$ found in \cite{kroon}.\\
(c) The same trivial reason might also explain why the large region 
with ${\bf Z}_3$ symmetry at $25.74 \leq Q \leq 25.84$ has 
been missed in \cite{kroon}. But there is also the more interesting 
possibility that it was missed since the attractor there is nearly 
(but not quite!) symmetric under the ``rotation" $g=\zeta$. As we 
can see from fig.5, we have to go to rather small distances to see 
unambiguously that $r_\zeta(\epsilon)\neq 1$, and this might easily 
be overlooked when using detectives.\\
(d) For most points in the interval $25.18 \leq Q \leq 25.32$ we 
find the intermittent behavior illutrated in fig.11 which suggests 
that the actual symmetry is the full ${\bf D}_6$. In contrast, 
the authors of \cite{kroon} quote ${\bf Z}_6$ for $Q=25.2$ and 
$Q=25.3$, which is the symmetry observed during the (short) times 
when the curve in fig.11 is below 1 (see also the discussion in 
\cite{kroon} concerning $Q=25.55$).\\
(e) For $25.35 \leq Q \leq 25.52$ we observe a large window with 
very clean ${\bf Z}_2$ symmetry, while \cite{kroon} quotes ${\bf Z}_6$ 
also for $Q=25.4$. The reason for this discrepancy is not clear to 
us. There is a ${\bf Z}_6$-symmetric transient, but its life time 
$t_{trans}\approx 100$ should be too small to have created any 
confusion.

\section{Conclusion}

We have shown that \ccs s are very efficient for studying symmetries 
of strange attractors. They are numerically more cumbersome than the 
``detectives" used in \cite{kroon,barany,dellnitz,chossat}, 
but they can be used for smaller data sets and they give much 
more detailed information. In particular, they allow easily to 
estimate the amount by which a symmetry is violated. 

We tested our method on a model with ${\bf D}_6$ symmetry studied 
by means of detectives in \cite{kroon}. We succeeded to obtain 
unique symmetry attributions even for control parameter values 
where the authors of \cite{kroon} failed in spite of much 
larger data sets. Part of the reason for this is due to rather 
trivial modifications which are independent of the method used for 
symmetry detection. These include discarding very long transient 
and using decimated time series (eventually with irregular time 
gaps, in order not to be influenced by commensurability effects 
between the sampling frequency and some periodicity of the 
attractor). But even taking this into account, it seems that 
our method makes more complete use of the information contained 
in a given time series.

Just as the authors of \cite{kroon}, we did not bother with formal 
statstical error estimates. Nevertheless, and in spite of the small 
samples used (5000 points), we feel that purely statistical errors 
were not a serious problem in the present application. Whenever we 
found an ambiguity, it was due to systematic effects (long transients 
and/or huge intermittency time scales) which cannot be estimated 
reliably by any method. At least, our method issues a warning in such 
a case in form of near-constant cross correlation ratios.

In addition, of course, the estimation of correlation sums 
allows also to study scaling properties of the attractor. Usually, 
correlation sums are computed only for that purpose, i.e. in 
order to estimate attractor dimensions. We have not yet mentioned 
them here on purpose, to stress that correlation sums contain much 
more useful information than just dimensions. Nevertheless, we 
might take the chance here to say that the attractor dimensions 
indicated by figs. 1-4 are between 2 and 5.

\bigskip

For discussions we thank H.-M. Br\"oker, H. Frauenkron, F. Krause, T. 
Sch\"urmann and M. Dellnitz. This work was partly supported by the 
Deutsche Forschungsgemeinschaft within SFB 237 and Graduiertenkolleg 
``Feldtheoretische und numerische Methoden in der Statistischen 
und Elementarteilchenphysik".

\eject

\end{document}